\documentclass[abstract=true]{scrartcl}

\usepackage[utf8]{inputenc}
\usepackage[T1]{fontenc}
\usepackage{amsmath,amssymb}
\usepackage{graphicx}
\usepackage{braket}
\usepackage{xcolor}
\usepackage{authblk}
\usepackage{hyperref}

\newcommand{\eV}{\textnormal{\,eV}}
\newcommand{\keV}{\textnormal{\,keV}}
\newcommand{\MeV}{\textnormal{\,MeV}}
\newcommand{\GeV}{\textnormal{\,GeV}}
\newcommand{\TeV}{\textnormal{\,TeV}}
\newcommand{\MU}{M_\textnormal{GUT}}
\newcommand{\MP}{M_\textnormal{Pl}}

\DeclareMathOperator{\Tr}{Tr}
\DeclareMathOperator{\diag}{diag}

\newcommand{\Eqref}[1]{eq.~\eqref{#1}}
\newcommand{\Figref}[1]{fig.~\ref{#1}}
\newcommand{\Tabref}[1]{tab.~\ref{#1}}
\newcommand{\Secref}[1]{sec.~\ref{#1}}

\begin{document}

\titlehead{\begin{flushright} FTPI-MINN-19/23\\
UMN-TH-3832/19 \end{flushright}}

\title{The Price of Tiny Kinetic Mixing}

\renewcommand\Affilfont{\itshape\small}
\author[1]{\large Tony Gherghetta\thanks{tgher@umn.edu}}
\author[2]{Jörn Kersten\thanks{joern.kersten@uib.no}}
\author[1,3]{Keith Olive\thanks{olive@umn.edu}}
\author[1,3,4]{Maxim Pospelov\thanks{mpospelov@perimeterinstitute.ca}}
\affil[1]{School of Physics and Astronomy, University of Minnesota,
Minneapolis, MN 55455, USA}
\affil[2]{University of Bergen, Institute for Physics and Technology,
Postboks~7803, 5020 Bergen, Norway}
\affil[3]{William I.~Fine Theoretical Physics Institute, School of
Physics and Astronomy, University of Minnesota, Minneapolis, MN 55455, USA}
\affil[4]{Perimeter Institute for Theoretical Physics, Waterloo, ON N2J 2W9, Canada}

\setkomafont{date}{\normalsize}
\date{November 1, 2019}
 
\maketitle

\begin{abstract}
\noindent
We consider both ``bottom-up'' and
``top-down'' approaches to the origin of gauge kinetic mixing. 
We focus on the possibilities for obtaining kinetic mixings $\epsilon$ which are consistent with experimental constraints and are much smaller than the naive estimates ($\epsilon \sim 10^{-2} - 10^{-1}$) at the one-loop level. In the bottom-up approach, we consider the possible suppression from multi-loop processes. Indeed we argue that kinetic mixing through gravity alone, requires at least six loops and could be as large as $\sim 10^{-13}$. In the top-down approach we consider embedding the Standard Model and a $U(1)_X$ in a single grand-unified gauge group as well as the mixing between Abelian and non-Abelian gauge sectors.  
\end{abstract}

\section{Introduction}

While we can be quite certain of the existence of dark matter (DM),
we can with equal certainty claim that we have no idea as
to the nature or identity of the dark matter, as it pertains to its connection
to fundamental particle physics. This is not because of the lack of options, 
but rather due to a great multitude of possibilities for DM\@. 
Some well-motivated weak-scale candidates 
such as a fourth-generation heavy neutral
lepton \cite{gllss}, have long been excluded by the width of the $Z$ gauge boson \cite{ALEPH:2005ab} and direct detection experiments \cite{Ahlen,Caldwell,Beck}.
However, most DM models have been only partially constrained, 
rather than outright excluded. 
This includes supersymmetric DM candidates \cite{Gold,ehnos} that 
so far have been absent in LHC searches \cite{LHC1,LHC2,LHC3,LHC4}, 
and in direct detection experiments \cite{Akerib:2016vxi,Cui:2017nnn,Aprile:2018dbl}. 
Ultralight DM, including axions \cite{pww,ab,df}, could be another generic option, but no positive evidence for DM of this kind has emerged thus far either. 

Given the lack of a clear top-down preference for DM, an alternative
approach has been pursued in recent years, that consists of
investigating simple UV-complete theories of particle DM\@. This
approach has led to the concept of ``dark sectors'', which include not only the DM particles but also possible force carriers that allow the DM to interact with itself and/or with the Standard Model (SM) \cite{Jaeckel:2010ni,Jaeckel:2013ija,Alexander:2016aln}. 
Constrained only by the fundamental principles of gauge invariance, anomaly cancellation etc., such an approach leaves many possibilities open, and usually does not predict the strength of the interaction from first principles. This can be contrasted with the framework provided by supersymmetry, 
where the interaction strength can often be fixed from first principles. 
Indeed, one of the attributes of supersymmetry as an
extension of the SM is the specific nature of the interactions
between the new particles and SM particles, as they are all related to gauge
or Yukawa interactions using known supersymmetric transformations. Although very difficult to detect, even the gravitino interactions with matter can be predicted.

In the dark sector approach, the interaction of DM with the
SM can occur through one (or several) portals. 
For the classification and current experimental constraints, see
{\em e.g.}~the recent reviews \cite{Battaglieri:2017aum,Beacham:2019nyx}. 
The phenomenology of new Abelian gauge bosons, as possible mediators of DM-SM interactions, has been extensively studied in the literature \cite{Alves:2013tqa,Lebedev:2014bba,Arcadi:2013qia}. Being electrically neutral, such new gauge bosons may exist in  a wide mass range, from the sub-eV energy scale to the weak scale and beyond. The gauge boson mass may be due to 
some spontaneous breaking of a dark gauge group, or in the Abelian case may be given by a 
Stückelberg term in the Lagrangian. 

\begin{figure}[ht]
\centering
\includegraphics{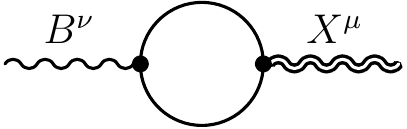}
\caption{A Feynman diagram depicting the generation of kinetic mixing at the 1-loop level.}
\label{fig:Classic1Loop}
\end{figure}

The most natural way of coupling the SM fields to the dark sector 
is via the so-called kinetic mixing operator. 
Kinetic mixing occurs whenever a term such as
\begin{equation}
    \mathcal{L} \supset \epsilon \,\frac{1}{2} F^{\mu\nu} X_{\mu\nu}~,
\end{equation}
appears in the Lagrangian where $\epsilon$ is a dimensionless parameter. Here $F_{\mu\nu} = \partial_\mu A_\nu -
\partial_\nu A_\mu $ is the electromagnetic field strength
which is related to the $U(1)_Y$ hypercharge field strength $B^{\mu\nu}$ via $\cos\theta_W$ where $\theta_W$ is the weak mixing angle,
and $X_{\mu\nu}= \partial_\mu X_\nu -
\partial_\nu X_\mu $ is the field strength for a hidden sector $U(1)_X$
gauge boson, $X_\mu$.
Assuming that the kinetic mixing vanishes at a high scale and there are fields charged under both $U(1)$'s, the Feynman diagram in
\Figref{fig:Classic1Loop} yields the well-known result
\cite{Holdom:1985ag,Cheung:2009qd}
\begin{equation} \label{eq:Standard1L}
	\epsilon = -\frac{g' g_X}{16\pi^2} \sum_i Y_i q_i \ln\frac{M_i^2}{\mu^2}\,,
\end{equation}
for kinetic mixing with $U(1)_Y$ at the one-loop level. 
Here,
$g'$ and $g_X$ are the gauge couplings of the two $U(1)$'s,  $Y_i$ and $q_i$
are the respective charges of the fields in the loop with mass $M_i$, and $\mu$ is a renormalization scale.
In the absence of precise cancellations, this leads to an estimate of
$\epsilon \sim (10^{-2}-10^{-1})\times g_X$, depending on the exact field content of particles running in the loop, and the scale separation in the logarithm.
The kinetic mixing with the photon is obtained by multiplying $\epsilon$
by $\cos\theta_W$, which does not change the order of magnitude estimate for the mixing.
Consequently, to obtain the small amount of mixing required by 
experimental limits \cite{Jaeckel:2010ni,Jaeckel:2013ija,Alexander:2016aln}, we need either a very small gauge coupling for the new
$U(1)_X$ or an alternative mechanism which generates kinetic mixing.

\begin{figure}[ht]
\centering
\includegraphics{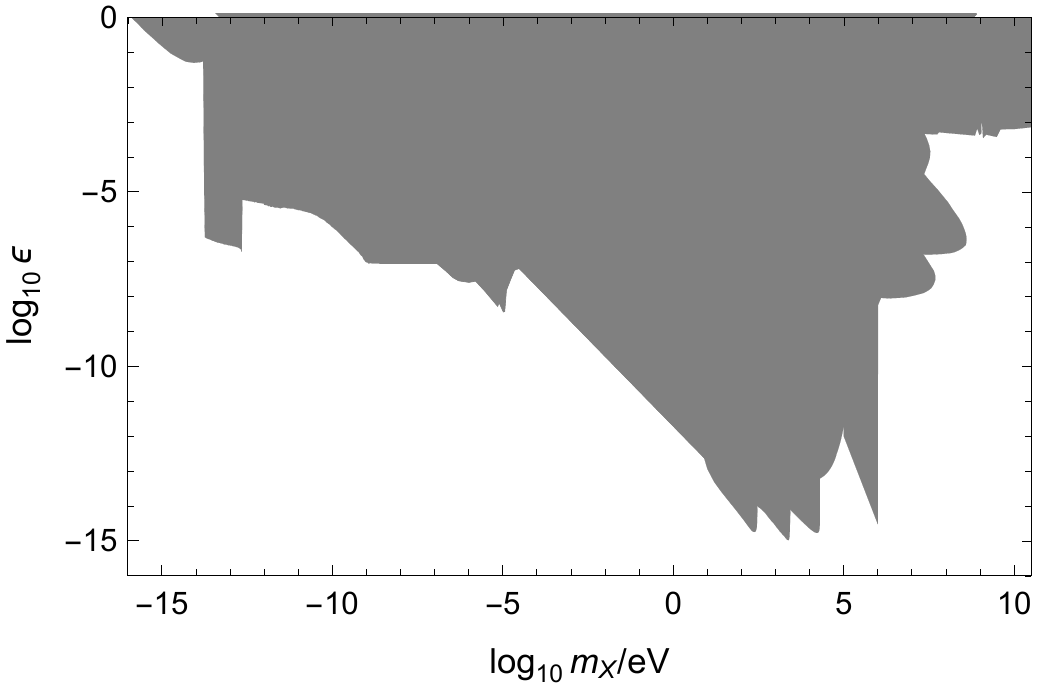}
\caption{A summary of the experimental bounds on kinetic mixing, showing the
strongest available bound for each dark photon mass $m_X$. Adopted from
\cite{Redondo:2015iea}. Not shown are the additional ``islands'' of CMB- and 
BBN-excluded regions extending down to $\epsilon \sim 10^{-18}$ for $m_X$ in the MeV-range \cite{Fradette:2014sza}.}
\label{fig:Bounds}
\end{figure}

In \Figref{fig:Bounds}, we show the strongest bounds on $\epsilon$ as a function of the dark photon mass. These limits come from a variety of sources which include
the magnetic field of Jupiter~\cite{1999TJPh...23..943P},
the Cosmic Microwave Background \cite{Jaeckel:2008fi,Mirizzi:2009iz},
searches for deviations from Coulomb's law~\cite{Williams:1971ms},
the CERN Resonant WISP Search (CROWS) \cite{Betz:2013dza,Graham:2014sha},
extra energy loss of stars \cite{An:2013yfc,Redondo:2013lna,Vinyoles:2015aba},
effects of dark photon decay on cosmology \cite{Redondo:2008ec},
SN1987A \cite{Bjorken:2009mm},
as well as
fixed target experiments and searches for dilepton resonances \cite{Beacham:2019nyx}.

We see that the limits on the kinetic mixing parameter at the sub-GeV
scale are below the value found at one loop, which is thus too large for many phenomenological applications.
Notable examples of constraints on $\epsilon$ include the above mentioned astrophysical constraints
on a eV-to-$100\keV$ mass $X$ boson, where the constraint on $\epsilon$
can be as tight as $10^{-15}$ \cite{An:2013yfc,Redondo:2013lna}. In
addition, DM masses in the range of 10 to $100\MeV$ and $X$-mediated freeze-out 
often require
values for the kinetic mixing between $10^{-5}$ and $10^{-3}$ 
\cite{Batell:2009di,Izaguirre:2013uxa}, which are also in 
tension with the one-loop estimate.
Also note that \Figref{fig:Bounds} refers to the limits on $\epsilon$ when the $X$ gauge boson has a Stückelberg-type mass.
A dark Higgs origin for~$m_X$ results in a stronger bound in the entire 
range $m_X \lesssim 10\keV$, where the combination $\epsilon\times g_X$
is limited to $\lesssim 10^{-14}$ from the energy loss by dark Higgs
emission in stars, in particular red giants \cite{Ahlers:2008qc}.

While a phenomenological (or ``bottom-up'') approach does not 
single out any particular value for $g_X$ and $\epsilon$, significant restrictions 
on their value may come from a theoretical requirement of gauge coupling unification. 
While there are different ways of  embedding the SM in a grand unified theory (GUT), there are 
few attempts for augmenting the SM with a new ``dark'' $U(1)$ gauge group.
One of the questions we wish to address in this paper is the level of kinetic mixing
any new gauge interaction may have with the SM (the photon in particular), 
in the context of a GUT. 

If the SM is unified into a GUT, the hidden gauge bosons
may be embedded at some scale into a GUT gauge group larger than $SU(5)$.
If not, kinetic mixing with the unified field strength will require
the presence of effective operators coupling the adjoint representation of the GUT with
the hidden sector. We will discuss both of these possibilities
with a view of estimating how large or small kinetic mixing may be.

The GUT-based approach, interpreted naively as $\alpha_X \sim \alpha_{\text{SM}}$,
may not be inevitable in the top-down approach. 
Indeed, in the literature, LARGE volume string compactifications
have been pointed out as a way to obtain very
small gauge couplings $g \sim 10^{-4}$
(or $\alpha \sim 10^{-9}$)~\cite{Burgess:2008ri} and
tiny kinetic mixing via \Eqref{eq:Standard1L} \cite{Cicoli:2011yh}.
Alternatively, in string theory extra $U(1)$'s are ubiquitous either from the closed string sector~\cite{Dienes:1996zr} (including e.g.~RR photons~\cite{Camara:2011jg}) or open string hidden sectors~\cite{Bianchi:2013gka}, and these can mix with the visible sector. 

Independent of any GUT, we explore the phenomenological ranges of kinetic mixing that may receive additional suppression from multi-loop 
mechanisms. Surprisingly, kinetic mixing may also occur through
purely gravitational interactions, provided that there is a source of 
charge symmetry breaking in the dark sector. We argue that this particular type of mixing through
gravity requires at least six loops. Although heavily suppressed by the gravitational coupling and loop factors, a non-negligible mixing of order $10^{-13}$ is possible with a Planck scale cutoff.
Furthermore we argue that this is the minimum kinetic mixing in any theory with hidden gauge interactions and charge symmetry breaking in the dark sector.

The outline of this paper is as follows: 
We begin with a survey of phenomenological (bottom-up) approaches to kinetic mixing, including possible multi-loop generation mechanisms. 
In particular we discuss mechanisms via graviton exchange, and point out the conditions needed to generate this particular type of kinetic mixing.
In \Secref{single}, we survey the various 
top-down possibilities for grand unification which includes the hidden sector.
The generation of effective operators that mix an extra $U(1)_X$ with a
SM GUT is discussed in \Secref{sec:NonAbelian}.
Our conclusions are given in \Secref{summary}.

\section{Phenomenological (Bottom-Up) Approaches}

In this section, we consider some ideas for generating kinetic mixing using a bottom-up approach, demonstrating a wide variety of possibilities. However before we do that, some general comments based on symmetry arguments are in order. Consider the schematic Lagrangian
\begin{equation}
    {\cal L}= {\cal L}_A + {\cal L}_X + {\cal L}_{\text{int}}(X,A) \,,
\end{equation}
that includes two ``separate'' Lagrangians, ${\cal L}_{A,X}$ which contain
kinetic terms for gauge bosons and their interaction with currents built from matter fields, ${\cal L}_A = -\frac{1}{4} F_{\mu\nu}^2 - A^\mu J_\mu^{(A)} +\dots$
Here $J_\mu^{(A)}$ is the current of particles charged only under a $U(1)_A$ gauge group.
The interaction Lagrangian between the two sectors can include kinetic mixing as well as 
other generic forms of interactions between the fields charged under
$U(1)_A$ and $U(1)_X$.  One can introduce two separate charge conjugation symmetries, 
${\cal C}_A$ and ${\cal C}_X$ that act on the 
fields as ${\cal C}_A(A) =-A$, ${\cal C}_X(X) =-X$.
The operator $F^{\mu\nu}X_{\mu\nu}$ is obviously odd under these separate charge symmetry transformations. Notice that if $X$ is massless and there is no 
matter charged under $X$, the kinetic mixing operator can be removed by a
$(A,X)$ field redefinition. In this case, even in the presence of the kinetic 
mixing operator, one can define two independently conserved charge conjugation 
symmetries. However the introduction of a mass term, $m_X^2X_\mu^2$, makes $\epsilon$
observable, so that it is the $\epsilon \times m_X^2$ parameter that breaks 
two charge symmetries down to one common ${\cal C}$.

If ${\cal C}_A$ and 
${\cal C}_X$ are separately good symmetries of the full Lagrangian, 
then kinetic mixing cannot be induced at any perturbative order
\cite{Dienes:1996zr,Garny:2018grs}.
In order to 
generate kinetic mixing, the individual charge symmetries must be broken, either 
completely or down to a common charge symmetry. 
For example, if both 
${\cal L}_A$ and ${\cal L}_X$ are QED-like, then 
${\cal C}_A({\cal L}_A) ={\cal L}_A$ and ${\cal C}_X({\cal L}_X) ={\cal L}_X$. If in addition the interaction 
term ${\cal L}_{\text{int}}$ is also invariant under {\em separate} charge symmetries, then the kinetic mixing term cannot be generated. 

As an explicit example, consider  two 
scalar QED theories with one field $\phi$ charged under $A$, 
and another field $\chi$ charged under $X$ with an interaction Lagrangian 
in the form of a scalar portal, ${\cal L}_{\text{int}}=
-\lambda (\phi^\dagger \phi)(\chi^\dagger \chi)$. In such a theory, the full Lagrangian 
${\cal L}$ is invariant under separate charge conjugation symmetries, 
and therefore kinetic mixing will never develop at any perturbative order because at least one of the ${\cal C}$ symmetries would need to be violated, either in ${\cal L}_{A,X}$ or in ${\cal L}_{\text{int}}$. 

The one-loop example from the previous section demonstrates that commonly 
charged matter does indeed break individual charge conjugation symmetries 
down to a common ${\cal C}$-symmetry. In other words, 
matter interactions with both  gauge bosons, {\em e.g.} $\bar \psi \gamma_\mu D^\mu_{AX}\psi $, where $D^\mu_{AX}$ is the covariant derivative with respect to the $A$ and $X$ fields, cannot be made separately ${\cal C}_A$ and 
${\cal C}_X$ symmetric. This interaction is of course invariant under a usual
charge conjugation symmetry: ${\cal C}(\bar\psi\gamma_\mu\psi) = - \bar\psi\gamma_\mu\psi$, under which both fields are transformed, ${\cal C}(X) = -X$, and ${\cal C}(A) = -A$.

Moreover, the charge conjugation symmetry is indeed maximally violated in the SM, 
as is parity, due to a drastic asymmetry in the charge assignments between the left- and right-handed fields. However this does not mean that kinetic mixing will be induced 
for any ``dark'' gauge boson $X$, as ${\cal C}_{X}$ must also be broken. 
Therefore the most crucial assumptions affecting the kinetic mixing depend on the structure of the dark $X$-sector (QED-like or chiral, SM-like)
and the presence or absence of commonly charged matter fields.
In all the examples considered below, we will assume that the separate ${\cal C}_X$ symmetry is violated.

\subsection{Gauge-Mediated Kinetic Mixing}
\label{minimal}

We begin with the one-loop estimate of Holdom, \Eqref{eq:Standard1L}, and
``work our way down''
in~$\epsilon$ by pursuing different choices of $X$ interactions. What are the generic ways of 
making the kinetic mixing $\epsilon$ smaller without assuming the gauge couplings are tiny?

In the bottom-up picture, we do not have any information about the
tree-level value of $\epsilon$ at very high energies, which is determined
by unknown UV physics.  We are therefore restricted to determining the
radiative corrections in the low-energy theory.  These can be viewed
either as the result of the running of $\epsilon$ from high to low
energy or as loop corrections evaluated directly at the low-energy scale
relevant for observations.  The results will usually depend on an
unphysical renormalization scale $\mu$, as in \Eqref{eq:Standard1L}, for
example.  As long as we do not specify the precise observable sensitive
to $\epsilon$, it is not obvious which value to choose for $\mu$.
However, since $\mu$ only appears logarithmically, this does not
introduce an uncertainty of more than an order of magnitude, which is
sufficient for our purposes.

We will consider the value of the lowest-order non-zero correction to
$\epsilon$ as a generic lower limit.  Of course, smaller values can be
obtained if there is a cancellation between a non-zero tree-level value
and radiative corrections.%
\footnote{Such fine-tuning can have the upside of
an interesting cosmology \cite{Banerjee:2019asa}.}

One obvious possibility for suppressing $\epsilon$ is to introduce several particles in 
the commonly charged sector in such a way that the sum in \Eqref{eq:Standard1L}
is small.
If, for example, there are two heavy matter fields, $\psi$ and $\chi$, with the same 
charges under one gauge group and opposite charges under the other, then
the kinetic mixing parameter is suppressed. Indeed, at a loop momentum scale much above the 
particle masses, the sum gives zero, and only \emph{threshold} effects due to 
$m_{\psi,\chi}$ give a nonzero result. Thus, in this case
we will have $\sum_{i=\psi,\chi}Y_i q_i \ln(M_i^2/\mu^2)$ simplifying to 
$Y_\psi q_\psi\ln(M_\psi^2/M_\chi^2)$ (or more precisely to a difference of polarization 
diagrams for $\chi$ and $\psi$).
In the limit of degenerate masses, the logarithm can be 
very small, approximately $\Delta M^2/M^2$, where $M$ is the common mass scale and $\Delta M$ is the mass splitting. Such a mass degeneracy could result from an underlying GUT 
symmetry, as further discussed in \Secref{single}.  Similar effects are also found in string theory, and result from an underlying mass degeneracy in the string spectrum~\cite{Dienes:1996zr}.

With the exception of matter fields with degenerate masses, kinetic mixing 
generated at  
one-loop is generically too large for the phenomenological applications discussed in the introduction.  
This suggests trying to realize the suppression of 
$\epsilon$  by devising a multi-loop generation mechanism. 
A known example is the mirror-symmetric twin Higgs model, where kinetic
mixing is at least four-loop-suppressed, leading to
$\epsilon \sim 10^{-13} - 10^{-10}$ \cite{Chacko:2005pe,Koren:2019iuv}.

\begin{figure}[ht]
\centering
\includegraphics{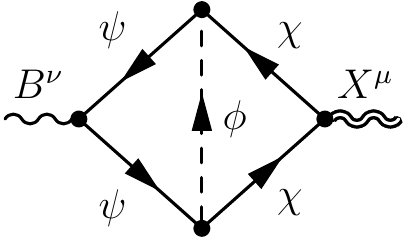}
\caption{A Feynman diagram depicting the generation of kinetic mixing at the 2-loop level.}
\label{fig:Generic2Loop}
\end{figure}

We begin with two loops, and it turns out that  
it is not entirely trivial to find a working example.
Consider the generic two-loop diagram in \Figref{fig:Generic2Loop}.  If
we choose $U(1)_Y \times U(1)_X$ charges $(q,0)$ for $\psi$ and $(0,q)$
for $\chi$, we obtain kinetic mixing if $\phi$ has charges $(q,-q)$,
while the one-loop diagram of \Figref{fig:Classic1Loop} with $\psi$ or
$\chi$ in the loop cannot contribute.  However, the analogous one-loop
diagram with $\phi$ in the loop \emph{does} contribute and will lead us
back to the estimate \eqref{eq:Standard1L}.

\begin{figure}[ht]
$\vcenter{\hbox{\includegraphics{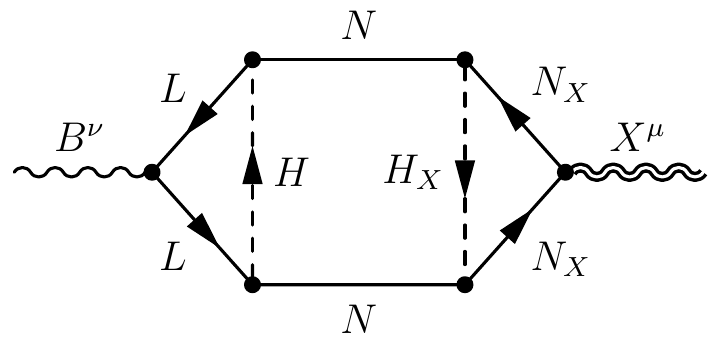}}} \hfill
 \vcenter{\hbox{\includegraphics{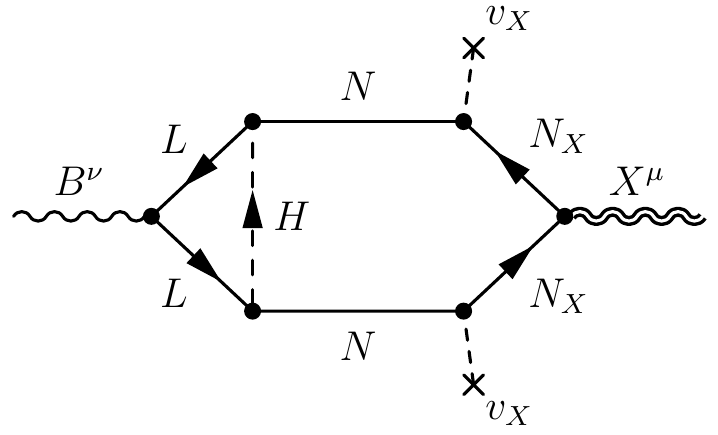}}}$
\caption{The Feynman diagrams depicting the generation of kinetic mixing in a neutrino portal model.}
\label{fig:Neutrino3Loop}
\end{figure}

A working example can be obtained at the three-loop level by using the
neutrino portal between active (SM) and sterile (SM-singlet) neutrinos.
We consider a $U(1)_X$ gauge boson that couples only to the sterile neutrino sector.
In addition to the ``standard'' Yukawa interaction $y_N LHN_i$ (with Yukawa coupling $y_N$) that couples heavy
singlet neutrinos $N_i$, with Majorana mass $m_N$, to the SM Higgs $H$ and lepton doublet $L$, we introduce the $y_X N_i H_X N_X$ portal (with Yukawa coupling $y_X$) that further
couples $N_i$ to a Higgs field $H_X$ and a fermion $N_X$ charged under 
$U(1)_X$ \cite{Pospelov:2011ha}.
The typical mass hierarchy is $m_N \gg m_W \gg m_X > m_{N_X}$.\footnote{Note that at least two $N_i$ are needed to avoid a massless state.}
Kinetic mixing will be induced as shown on the left in \Figref{fig:Neutrino3Loop}, and we estimate
\begin{equation}
\epsilon \sim
\frac{y_N^2 y_X^2 g_X g'}{(16\pi^2)^3} \ln\frac{\mu^2}{m_N^2} \sim
10^{-7} \, (y_N y_X)^2 \,,
\end{equation}
assuming the log factor is of order one. 

By cutting the internal $H_X$ line, we can form a dimension-six operator, $B^{\mu\nu}X_{\mu\nu}H_X^\dagger H_X$ and after
replacing both $H_X$'s with the dark Higgs vacuum expectation value (vev), $v_X$, we obtain a two-loop
diagram shown on the right in \Figref{fig:Neutrino3Loop}, which gives a contribution of similar size, depending on
parameter values.
It is important to note that the result is now proportional to the Yukawa couplings $y_N$ and $y_X$. Therefore, the size of
the kinetic mixing can be dialed to an almost arbitrarily small value, 
by choosing $y_N y_X$ to be very small
(although doing so, may cause other model dependent problems with multiple very light fermions).

\begin{figure}[ht]
\centering
\includegraphics{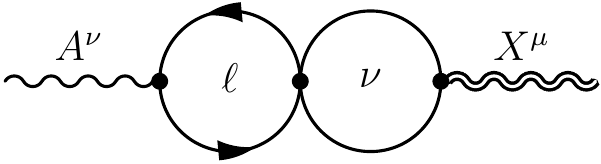}
\caption{The generation of kinetic mixing in the low-energy effective theory
arising from the neutrino portal model.  The particles in the loops are
a charged lepton $\ell$ and a neutrino mass eigenstate $\nu$, which is a
mixture of a SM neutrino, $N_X$, and $N$.}
\label{fig:Neutrinoeyeglass}
\end{figure}

In the effective theory valid below the electroweak scale, which
corresponds to the model considered in \cite{Dasgupta:2013zpn,Bringmann:2013vra}, the three-loop diagram in  \Figref{fig:Neutrino3Loop} can be reduced to the two-loop  diagram shown in \Figref{fig:Neutrinoeyeglass} with a four-Fermi vertex.
After the electroweak symmetry and $U(1)_X$ are broken, SM neutrinos 
mix with $N_i$ and $N_X$.  Although kinetic mixing with
the photon cannot be generated at one loop since there is no field with
both an electric and a $U(1)_X$ charge, it can instead arise from
\Figref{fig:Neutrinoeyeglass}.
A very rough estimate is
\begin{equation} \label{eq:Sterile2Loop}
\epsilon \sim
\frac{e g_X}{(16\pi^2)^2} G_F m_X^2 \theta^2 \sim
10^{-17} \left( \frac{m_X}{1\MeV} \right)^2
\left( \frac{\theta}{0.1} \right)^2 ,
\end{equation}
where
$\theta$ is the active-sterile neutrino mixing angle
(e.g., $\theta \sim y_N v/y_X v_X$ if the
masses of~$N_i$ are similar and $m_{N_X} \gg m_\nu$)
and we have assumed $g_X \sim e \sim 1$.

Next we discuss mechanisms that use not only charged matter but also
intermediate gauge bosons of an additional third group.  Consider a
bottom-up model with two sufficiently heavy vector-like fermions $\psi$
and $\chi$ as well as an additional gauge group $U(1)_M$ that is
spontaneously broken at a high scale.  The charge assignments are
specified in \Tabref{tab:4LoopParticles}.

\begin{table}[ht]
\centering
\begin{tabular}{ccccc}
\hline
& Mass & \multicolumn{3}{c}{Charge} \\
& & $U(1)_Y$ & $U(1)_M$ & $U(1)_X$ \\
\hline
$\psi$ & heavy & $1$ & $1$ & $0$ \\
$\chi$ & heavy & $0$ & $1$ & $1$ \\
$B^\mu$ & light & $0$ & $0$ & $0$ \\
$M^\mu$ & heavy & $0$ & $0$ & $0$ \\
$X^\mu$ & light & $0$ & $0$ & $0$ \\
\hline
\end{tabular}
\caption{The particle content of a bottom-up model yielding kinetic mixing
at the 4-loop level.  Here ``light" refers to mass scales at the electroweak scale and below, while
``heavy" refers to mass scales significantly above the weak scale.}
\label{tab:4LoopParticles}
\end{table}

\begin{figure}[ht]
\centering
\includegraphics{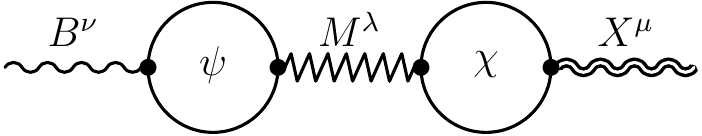}
\includegraphics{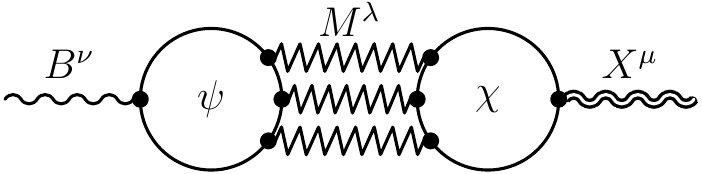}
\caption{The generation of kinetic mixing at the 4-loop level. The left Feynman diagram does not yield the correct operator and hence gives no contribution to
kinetic mixing.}
\label{fig:BottomUp4Loop}
\end{figure}

The two-loop diagram in \Figref{fig:BottomUp4Loop} is proportional to
$\Pi^{\nu\rho}_{YM}(k^2) \, D^{\rho\sigma}_M \Pi^{\sigma\mu}_{MX}(k^2)
\sim k^4/m_M^2$, where $\Pi$ denotes a self-energy contribution.
Consequently, this diagram leads to an operator containing derivatives
of $B^{\mu\nu}$ and $X^{\mu\nu}$ and thus does not contribute
to kinetic mixing.
The corresponding three-loop contribution with a second $U(1)_M$ gauge boson
vanishes due to Furry's theorem (diagrams containing a closed fermion
loop with an odd number of vertices do not contribute).
Consequently, the leading contribution to kinetic mixing stems from the
four-loop diagram in \Figref{fig:BottomUp4Loop}, which is of course
highly suppressed,
\begin{equation}
\epsilon \sim \frac{g^\prime g_X g_M^6}{\left( 16\pi^2 \right)^4} \sim 10^{-9} \,,
\end{equation}
where $g_M$ is the $U(1)_M$ gauge coupling.
A similar mechanism for generating kinetic mixing was discussed recently
in \cite{Dunsky:2019api}, with the intermediate gauge group
corresponding to a Yang-Mills field. 

\subsection{Gravity-Mediated Kinetic Mixing} \label{sec:Gravity}
So far we have considered the outcome for the kinetic mixing parameter $\epsilon$, when there exist matter fields 
commonly charged under both the SM and the dark $U(1)_X$ or another new gauge group. We have seen that there is considerable freedom 
in the choice of the mediation mechanism, and as a consequence, in the expected value of $\epsilon$. 

In this subsection, we would like to address the question of how gravitational interactions alone could result in a 
finite kinetic mixing parameter. We imagine a series of diagrams that join the SM and the $U(1)_X$ sector by gravitational 
interactions, {\em i.e.} loops of gravitons. 
The size of such diagrams is controlled by some $n$-th power of the gravitational constant, $G_N \equiv \MP^{-2}$. 
The dimensionless nature of $\epsilon$ tells us that such diagrams may indeed be UV divergent, and one could expect
the result to scale as $\propto \Lambda_{\text{UV}}^{2n}/\MP^{2n}$. Since the UV cutoff, $\Lambda_{\text{UV}}$ could be comparable to the 
Planck mass $\MP$, the extreme smallness of the denominator can be mitigated by a larger numerator, rendering this to be a very 
UV-sensitive mechanism.

First we consider a case when the SM is supplemented by a non-interacting dark $U(1)_X$. 
While the charge conjugation symmetry  is broken in the SM, as discussed earlier, there is a separate charge conjugation 
symmetry, ${\cal C}_X$ in the dark sector, $X_{\mu\nu} \to - X_{\mu\nu}$ that leaves the action invariant (for instance, the dark sector could be QED-like). At the perturbative level this means that
any vertex between the gravitons and the $X$-boson will contain an even number of gauge fields, $X_\mu$.
Therefore, the perturbative result in this case is $\epsilon=0$. Since gravity is expected to preserve both discrete and gauge symmetries, we do not expect this conclusion to change even at a non-perturbative level. 

\begin{figure}[ht]
\centering
\includegraphics{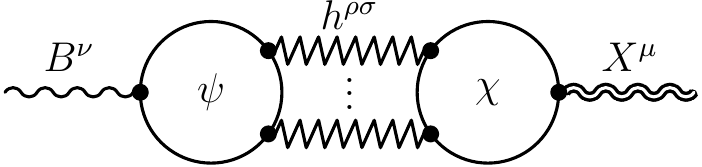}
\caption{The mediation of kinetic mixing via gravity, where the dots represent an unknown number of additional gravitons.}
\label{fig:GravitySchematic}
\end{figure}

If, on the other hand, there exists some matter content of the dark sector that results in a {\em separate} breaking of the dark 
charge conjugation symmetry, then there is a possibility of inducing non-zero kinetic mixing by means of gravity mediation. Consider, for example, 
a theory that contains a ``mirror'' SM-like sector, SM$'$, but no commonly charged fields 
under any of the SM and SM$'$ gauge groups. Schematically, the action of such a theory can be approximated by the sum of three terms,
\begin{equation}
{\cal S} = {\cal S}_{\text{SM}} + {\cal S}_{\text{SM}'} +{\cal S}_{\text{gravity}}
\,.
\end{equation}
Both SM and SM$'$ necessarily participate in gravitational interactions,
such that a diagram schematically shown in \Figref{fig:GravitySchematic} is always 
possible. The middle section of this diagram connecting two fermion loops in the SM and SM$'$ sectors contains 
an unknown number of gravitons, $h^{\rho\sigma}$.

It turns out that the minimum number of such intermediate gravitons is three. The best way of showing this is by cutting the 
diagram through the intermediate gravitons, and representing the left- and right-handed parts as effective operators composed of the 
$U(1)$ field strength and gravitationally gauge invariant operators. To be gauge invariant, these operators must be composed of the metric $g$ and gauge-invariant derivatives of the metric, {\em i.e.} the curvature ${\cal R}$:
\begin{equation}
F_{\mu\nu}\times {\cal O}^{\mu\nu}; ~~ {\cal O }= {\cal O}(g,~{\cal R} ) \,.
\end{equation}

It is easy to see that for one or two intermediate graviton exchanges the operator ${\cal O}^{\mu\nu}$ either does not exist 
or can be reduced to a total derivative, such that the operator $F_{\mu\nu} {\cal O}^{\mu\nu}$ would not lead to kinetic mixing. 
For one intermediate graviton all possible candidate structures for ${\cal O}^{\mu\nu}$ must contain at most one power of the curvature, such as 
$g^{\mu\nu};~{\cal R}^{\mu\nu};~ \nabla^\mu\nabla^\nu {\cal R}$ etc, where $\nabla^\mu$ is the gravitational covariant derivative. All of these structures are 
$\mu\leftrightarrow\nu$ symmetric, and give zero upon contraction with either $X_{\mu\nu}$ or $F_{\mu\nu}$. For two intermediate gravitons, we also find that the required ${\cal O}^{\mu\nu}$ tensors do not exist. The following candidate structures 
are explicitly symmetric under the interchange of  indices contracted with the $U(1)$ field strength $F_{\mu\nu}$: 
${\cal R}^{\mu\alpha\beta\nu}{\cal R}_{\alpha\beta}$, ${\cal R}^{\mu\alpha}{\cal R}^{\nu}_{\alpha}$.
Expressions that contain extra derivatives, such as ${\cal R}^{\mu\alpha}\nabla_{\alpha}\nabla^\nu {\cal R}$
and ${\cal R}^{\mu\alpha}\nabla^2 {\cal R}^{\nu}_{\alpha}$ can be simplified using integration by parts, and the result is either 
$\mu\leftrightarrow\nu$ symmetric, or contains $\nabla F$, and therefore does not lead to kinetic mixing. 

\begin{figure}[ht]
\centering
\includegraphics{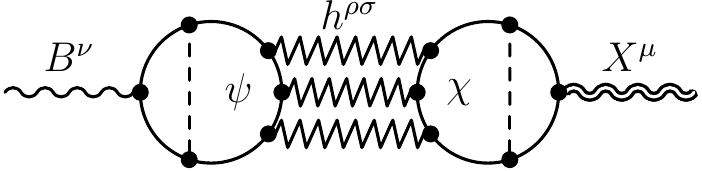}
\caption{The mediation of kinetic mixing via gravity showing the minimal three graviton exchange as well as the Higgs exchange inside the fermion loops to prevent a null result from gravitational anomalies.}
\label{fig:Gravity6Loop}
\end{figure}

Finally, at order ${\cal R}^3$, one can indeed find the required operators ${\cal O}^{\mu\nu}$ that do not vanish. 
These include structures like ${\cal R}^{\mu}_{\alpha}{\cal R}_{\lambda\rho}{\cal R}^{\nu\lambda\rho\alpha}$ and 
many other possible terms with derivatives. Such operators would generically lead to three graviton two-loop
exchanges  generating $\epsilon$. 
Moreover, the absence of a gravitational anomaly means that the sum of the respective  
hypercharges of all fermions in the SM and SM$'$ is zero. Therefore to avoid a null result the matter loops contain not only a fermionic loop, but also require an exchange by for example, the Higgs and
Higgs$'$ fields inside the fermionic loops, as shown in
\Figref{fig:Gravity6Loop}, so that $\Tr (Y_i y_i^2)\neq 0$, where $Y_i$ are the $U(1)$ charges and $y_i$ are the Yukawa couplings.

This raises the loop count to $6$, and we have the following extremely crude estimate: 
\begin{equation}
B_{\mu\nu} X_{\alpha\beta} \langle {\cal O}^{\mu\nu}  {\cal O}^{\alpha\beta}\rangle~\to ~
\epsilon_{\text{grav}} \propto g^\prime g_X y_t^2y_X^2\left( \frac{1}{16\pi^2}\right)^6 \times \frac{\Lambda_{\text{UV}}^6}{\MP^6} \,,
\label{eq:epsgrav}
\end{equation}
where $g^\prime(g_X)$ are the $U(1) (U(1)_X)$ gauge couplings, $y_t$ is the top Yukawa coupling, and $y_X$ is the Yukawa coupling in SM$^\prime$.
In this expression, $\langle \dots \rangle$ stands for the result of the gravitational loop mediation of the 
${\cal R}$-containing operators.

If $\Lambda_{\text{UV}}$ is of the same order as the Planck mass, the gravitationally-induced 
kinetic mixing estimated in (\ref{eq:epsgrav}) could be as large as $\epsilon_{\text{grav}} \sim 10^{-13}$. Interestingly, probing such a small kinetic mixing
observationally is not out of the question: astrophysical probes of $\epsilon$ can be very sensitive, particularly if the dark sector mass scale is in the 
eV-to-keV range \cite{An:2014twa}. At the same time it is worth mentioning that in theories with a parametrically large number of 
species, {\em e.g.} when the SM is extended by ${\cal N}$-copies, one also expects that $\Lambda^2_{\text{UV}} \lesssim 
\MP^2 \times {\cal N}^{-1}$, and the proposals of Refs.~\cite{Dvali:2007hz,Arkani-Hamed:2016rle} 
are perhaps not challenged by this mechanism.

\subsection{Clockwork Mechanisms}
The clockwork mechanism was proposed to generate very small couplings in
the absence of small fundamental parameters~\cite{Giudice:2016yja}.
In its gauge theory implementation, we consider $N+1$ $U(1)$ symmetries
labeled by $i=0,\dots,N$ with corresponding gauge fields $A_\mu^i$ and
equal gauge couplings, $g$.  The gauge symmetry is broken to a single
$U(1)$ by the (equal) vevs
$\braket{\phi_j}=f/\sqrt{2}$ (for all $j=0,\dots,N-1$) of $N$ Higgs
fields $\phi_j$.  Each of these scalars has charges $(1,-q)$ under
$U(1)_j \times U(1)_{j+1}$ (and charge $0$ under the other groups).
Diagonalizing the mass matrix for the gauge bosons yields a massless
zero mode, the gauge boson of the unbroken $U(1) \equiv U(1)_X$.  Once
this group is broken as well, this field becomes the hidden photon.
If a field is charged only under $U(1)_N$, its coupling to
the hidden photon is exponentially suppressed,
$g_\text{eff} = \frac{N_0 \, g}{q^N}$,
where $N_0 \sim 1$ is a normalization factor.

\begin{figure}[ht]
\centering
\includegraphics{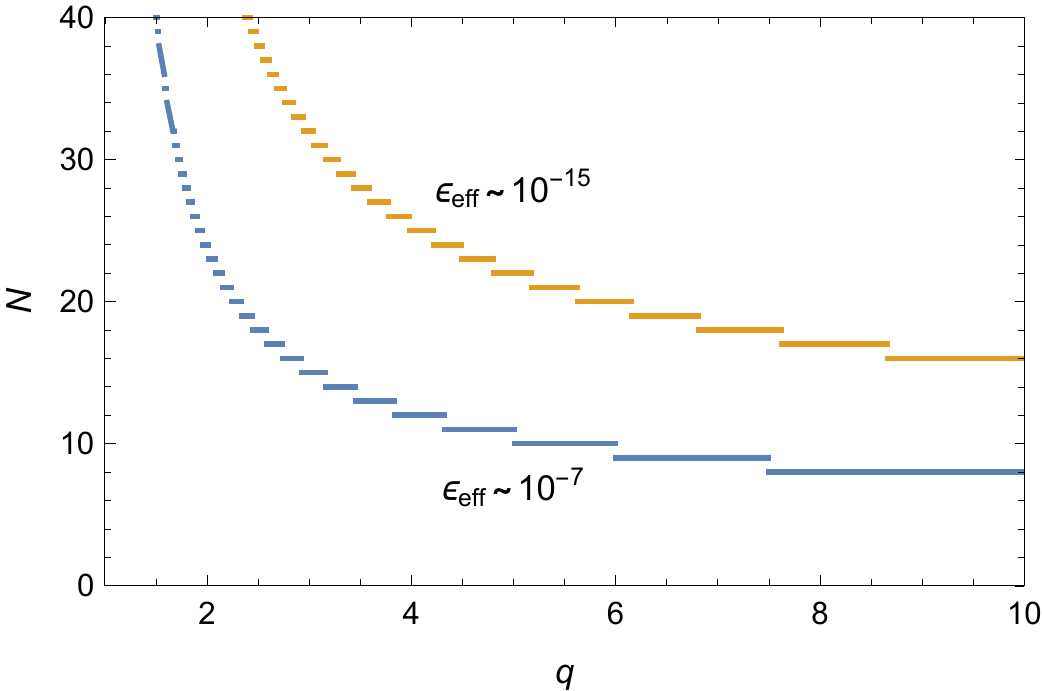}
\caption{The number of $U(1)$ gauge groups required to obtain the given kinetic
mixing parameter $\epsilon_\text{eff}$ via the clockwork mechanism as a function of the charge $q$.}
\label{fig:Clockwork}
\end{figure}

Likewise, if the $U(1)_Y$ gauge boson kinetically mixes only with
$A_\mu^N$, its kinetic mixing with the hidden photon is suppressed,
\begin{equation}
	\epsilon_\text{eff} = \frac{N_0 \, \epsilon}{q^N} \,.
\end{equation}
Thus, we can use the gauge clockwork mechanism to generate a tiny
kinetic mixing starting from $\epsilon \sim g \sim 1$.  The required
number of clock gears is given by
\begin{equation}
	N = \lceil \log_q\frac{N_0 \epsilon}{\epsilon_\text{eff}} \rceil ,
\end{equation}
where $\lceil x \rceil$ denotes the ceiling, i.e., the smallest integer
larger than $x$.  The result is shown in \Figref{fig:Clockwork} as a
function of $q$ for $N_0=1$ and two different values of $\epsilon_\text{eff}$.  For
example, $\epsilon_\text{eff} \sim 10^{-7}$ requires $N = 24$ for $q = 2$.

As quite a few $U(1)$'s are needed for a significant suppression, we
might consider the continuum limit $N \to \infty$, in which case the
clockwork mechanism becomes equivalent to a $5$-dimensional theory with localized bulk gauge bosons~\cite{Batell:2005wa}
and Higgs fields.  In this case the suppression factor becomes
$e^{-k L}$, where $L$ is the size of the extra dimension (for an orbifold $L=\pi R$ with $R$ the radius of the extra dimension), and $k$ is the
equivalent of $q$.

To summarize this section, we remark that the bottom-up approach 
leaves enough flexibility to cover a wide range of values of the mixing parameter $\epsilon$. 
Indeed, the one-loop result can be turned into a multi-loop generation
mechanism. Moreover, in certain examples given in this section, the kinetic mixing parameter vanishes if some corresponding Yukawa couplings vanish. 
Since Yukawa couplings are not necessarily fixed by unification, one could exploit 
some features of these mechanisms even within a GUT framework.

\section{Theoretical Top-Down Approaches}
The bottom-up approaches discussed so far have the disadvantage that
they can only provide lower limits on the size of kinetic mixing because
they do not contain mechanisms ensuring $\epsilon=0$ at tree level
(i.e., forbidding the term $F_{\mu\nu} X^{\mu\nu}$ in the original
Lagrangian).  In addition, these lower limits can be avoided by a
fine-tuned cancellation between a non-zero tree-level value and the loop
contributions considered above. 
We note that when the $U(1)_X$ gauge group is embedded in a GUT, we cannot assume a Stückelberg mass for the dark photon.  Instead, we must assume the presence of a dark Higgs of similar mass in which case the stronger limits
on $\epsilon$ discussed earlier apply. This will in addition require fine-tuning beyond that already 
needed for the doublet-triplet splitting in $SU(5)$,
in order to obtain a light $H_X$.
We now turn to top-down models where
the absence of kinetic mixing at a high-energy scale is guaranteed by a
symmetry.

\subsection{Embedding in a Single Group}
\label{single}

Let us first assume that both the SM gauge group and $U(1)_X$ are
embedded in the same group.  This implies that the rank of the group
is $5$ or larger.  In this case realistic symmetry
breaking patterns often lead to light states that are charged
under both $U(1)_Y$ and $U(1)_X$, and consequently to large kinetic 
mixing via \Figref{fig:Classic1Loop}.  However, for sufficiently large groups, it is possible
to construct counterexamples. 
In what follows, we consider progressively large gauge groups
and their symmetry breaking patterns and comment on
their suitability for generating
kinetic mixing. In particular,
we try to identify which group and field content could account
for mixing below the 1-loop estimate. 

\begin{description}
\item[$SO(10) \to SU(5) \times U(1)_X$\normalfont:]
The $SO(10)$ multiplet $\mathbf{16}$ decomposes into
$(\mathbf{1},-5) + (\mathbf{\overline{5}},3) + (\mathbf{10},-1)$
of $SU(5) \times U(1)_X$ \cite{Slansky:1981yr}, where the SM matter
fields are contained in the
$\mathbf{\overline{5}}$ and $\mathbf{10}$, which are both charged under
$U(1)_X$.  Equivalently, all SM matter is charged under $U(1)_{B-L}$,
which is related to $U(1)_X$ via $B-L = \frac{2}{5} Y - \frac{1}{5} X$.
Consequently, in this case we cannot obtain a kinetic mixing parameter
much below the Holdom estimate \eqref{eq:Standard1L}.

\item[$E_6 \to SO(10) \times U(1)_X$\normalfont:]
The $E_6$ multiplet $\mathbf{27}$
decomposes into $(\mathbf{1},-4) + (\mathbf{10},2) + (\mathbf{16},-1)$,
where the SM matter fields are in the $\mathbf{16}$ and
charged under $U(1)_X$.
Thus we would again obtain kinetic mixing at the 1-loop level.

\item[$E_6 \to SO(10) \times U(1)_A \to SU(5) \times U(1)_A \times U(1)_B$\normalfont:]
In this case we have two dark $U(1)$ groups at our disposal, which
allows us to choose $U(1)_X$ as a linear combination of $U(1)_A$ and
$U(1)_B$ such that either the $\overline{\mathbf{5}}$ or the
$\mathbf{10}$ of $SU(5)$ is uncharged under $U(1)_X$.  However, as these
multiplets stem from the same $\mathbf{16}$ of $SO(10)$, they have the
same $U(1)_A$ charge, whereas their $U(1)_B$ charges are different (see
first item).  As a consequence, one multiplet, either the $\overline{\mathbf{5}}$ or the
$\mathbf{10}$, unavoidably ends up with a
non-zero charge under both the SM $U(1)$ and $U(1)_X$.

\item[$E_7 \to E_6 \times U(1)_A \to SO(10) \times U(1)_A \times U(1)_B$\normalfont:]
We can again choose $U(1)_X$ as a linear combination of $U(1)_A$ and $U(1)_B$. In this case, 
we can ensure that the complete $\mathbf{16}$ of $SO(10)$ inside the
$\mathbf{27}$ of $E_6$ is uncharged.  Using LieART \cite{Feger:2012bs}
we find that the $E_7$ multiplet $\mathbf{56}$ decomposes into
$(\mathbf{1},3) + (\mathbf{1},-3) + (\mathbf{27},-1) +
 (\mathbf{\overline{27}},1)$.  Hence, the decomposition of the $\mathbf{27}$
under $SO(10) \times U(1)_A \times U(1)_B$ is
$(\mathbf{1},-1,-4) + (\mathbf{10},-1,2) + (\mathbf{16},-1,-1)$.
Consequently, the choice $X=A-B$ leads to a vanishing $U(1)_X$ charge
for all light \emph{matter} fields that arise from the 
$\mathbf{16}$.

However, the light Higgs belongs to a $\mathbf{10}$ of $SO(10)$, which
is usually assumed to arise from the same $E_6$ and $E_7$ multiplets as
the $\mathbf{16}$ containing the matter fields.  In this case, Higgs and
matter multiplets have the same $U(1)_A$ charge but different $U(1)_B$
charges, so their $U(1)_X$ charges cannot vanish simultaneously and we
again return to the Holdom estimate, this time due to a Higgs loop.  To
avoid this conclusion, we have to embed the $\mathbf{10}$ containing the
Higgs into a larger multiplet of $E_6$ in such a way that the ratio of $U(1)_A$
and $U(1)_B$ charges for this $\mathbf{10}$ is equal to the ratio for the matter
$\mathbf{16}$.  Using LieART we find that this is possible if the
$\mathbf{10}$ stems from the $\mathbf{133}$ of $E_7$ (which is the smallest representation beyond the $\mathbf{56}$).  This multiplet
decomposes into
$(\mathbf{1},0) + (\mathbf{27},2) + (\mathbf{\overline{27}},-2) +
 (\mathbf{78},0)$ of $E_6 \times U(1)_A$,
so the decomposition of the $\mathbf{27}$ is
$(\mathbf{1},2,-4) + (\mathbf{10},2,2) + (\mathbf{16},2,-1)$.  Now $X=A-B$
guarantees that the $U(1)_X$ charge vanishes for the $\mathbf{10}$ as well.

To summarize this example, we 
can ensure the vanishing of the 1-loop diagram for kinetic mixing in an $E_7$ GUT if we assume that (unlike more typical models of $E_6$ unification) the
$SO(10)$ Higgs multiplet (a $\mathbf{10}$) originates from a different $E_7$ multiplet than 
matter.  Matter fields sit inside the $\mathbf{16}$ of $SO(10)$, which sits inside a $\mathbf{27}$ of $E_6$, which sits inside the $\mathbf{56}$ of $E_7$. The $\mathbf{10}$ containing the Higgs also resides in a $\mathbf{27}$ of $E_6$, however, the latter originates from a $\mathbf{133}$ of $E_7$.
In this case, there are no light fields with non-zero charges under $U(1)_X$.

\item[$E_8 \to E_6 \times SU(3)$\normalfont:]
All SM fields can be assigned to the $E_8$ multiplet $\mathbf{248}$,
which decomposes into
$(\mathbf{1},\mathbf{8}) + (\mathbf{27},\mathbf{3}) +
 (\mathbf{\overline{27}},\overline{\mathbf{3}}) + (\mathbf{78},\mathbf{1})$,
where $(\mathbf{27},\mathbf{3})$ can accommodate the Higgs and matter
fields.  If we break $SU(3)$ to the $U(1)_X$ that is generated by the
diagonal $SU(3)$ generator $\lambda_3=\diag(1,-1,0)$,
there is an uncharged state in the triplet.  If in addition the other
two states obtain GUT-scale masses in the course of the symmetry
breaking, all light states remain uncharged under $U(1)_X$.
While this example, is simpler and all SM fields reside in a common $\mathbf{27}$ of $E_6$,
we are forced to a larger unification group and parent representation. In 
addition, in many $E_8$ unification models, the $SU(3)$ subgroup plays
the role of a (gauged) family symmetry
so that all three matter generations
reside in the $(\mathbf{27},\mathbf{3})$. That is not the case here, and
we must require a separate $\mathbf{248}$ for each generation. 

\item[$E_8 \to SU(5) \times SU(5)$\normalfont:]
We assume that the second $SU(5)$ contains $U(1)_X$ and we consider the
$E_8$ representations $\mathbf{248}$ and $\mathbf{3875}$. The options for
the SM matter multiplets are
$(\overline{\mathbf{5}},\mathbf{10})$,
$(\overline{\mathbf{5}},\mathbf{15})$,
$(\overline{\mathbf{5}},\mathbf{40})$,
$(\mathbf{10},\mathbf{5})$, and $(\mathbf{10},\mathbf{45})$.
Considering the decompositions of $\mathbf{5}$ and
$\mathbf{45}$ under $SU(5) \to SU(4) \times U(1)_X$ and
$SU(5) \to SU(3) \times SU(2) \times U(1)_X$, we find that there are
no states uncharged under $U(1)_X$.
However, if we do not restrict ourselves to maximal subgroups,
we can proceed as in the previous item and break $SU(5)$ to one of the
$U(1)$ subgroups under which for example, the multiplets $\mathbf{5}$ and
$\mathbf{10}$ contain uncharged states.
\end{description}

\noindent
While the next two examples are not specifically unified gauge groups, they have often been considered as UV extensions of the SM.

\begin{description}
\item[$SU(4) \times SU(2)_\textnormal{L} \times SU(2)_\textnormal{R}$\normalfont:]
The usual breaking to the SM by a
$(\mathbf{4},\mathbf{1},\mathbf{2})$ does not leave an extra $U(1)_X$.
If we use a $\mathbf{15}$ instead to break $SU(4) \to SU(3) \times U(1)_{X}$
(which yields the left-right symmetric model), $X = B-L$ and again all
SM matter fields are charged.

\item[$SU(3)_c \times SU(3)_\textnormal{L} \times SU(3)_\textnormal{R}$\normalfont:]
In the minimal trinification model \cite{Sayre:2006ma}, symmetry
breaking proceeds via two $(\mathbf{1},\mathbf{3},\overline{\mathbf{3}})$
scalars.  Individually, each vev breaks the gauge group to
$SU(3)_c \times SU(2)_\text{L} \times SU(2)_\text{R} \times U(1)$, but
the two scalars lead to different $SU(2)_\text{R} \times U(1)$ groups.
Consequently, in combination the vevs break $SU(3)^3$ directly to the
SM gauge group, leaving only a global $U(1)$.
Thus, a dark photon and kinetic mixing would require a
significantly modified scalar sector that leaves a local $U(1)$
unbroken.
\end{description}
If there are light fields charged under both $U(1)$'s, they are contained in complete GUT multiplets and then the diagram in \Figref{fig:Classic1Loop} vanishes for equal masses.
However, this does not decrease $\epsilon$ significantly at low
energies, where it will contain logarithms of particle masses, which are
not small for the SM particles (cf.~\Secref{minimal}).

In any case, heavy fields charged under both $U(1)_Y$ and $U(1)_X$ will
occur.  As they fill out complete GUT multiplets, their 
contribution to $\epsilon$ is sensitive to the mass splittings within
these multiplets caused by the GUT symmetry breaking.
If this leads to a mass splitting at tree level, we still obtain a
sizable value of $\epsilon$ via \Eqref{eq:Standard1L}.
However, if the mass degeneracy is only broken by renormalization group
running, kinetic mixing arises effectively at the two-loop level, so we
expect only $\epsilon \sim 10^{-6}-10^{-4}$ \cite{ArkaniHamed:2008qp}.
This is still too large to satisfy some experimental bounds, but an
additional suppression by one order of magnitude due to a small coupling
could be sufficient when $m_X \gtrsim 1$ MeV.

In summary, among commonly considered unified groups we find examples
without light fields charged under $U(1)_{Y,X}$ only for $E_7$ and $E_8$.  We do not
attempt to work out the model building details for these cases, which would
also have to address the emergence of chiral fields from the real
representations of $E_7$ and $E_8$ (as could, for example, arise from an orbifold compactification).

\subsection{Mixing between Non-Abelian and Abelian Sectors} \label{sec:NonAbelian}

If only one of the gauge groups involved is non-Abelian, the kinetic mixing
term $G^{\mu\nu} X_{\mu\nu}$ is forbidden by gauge invariance, since the
non-Abelian field strength $G^{\mu\nu}$ is not gauge-invariant.  Thus,
the diagram of \Figref{fig:Classic1Loop} vanishes even in the presence
of particles that are charged under both gauge groups.
However, we can realize kinetic mixing via effective operators involving
appropriate scalar representations, for example,
$\frac{1}{\Lambda} \Sigma \, G^{\mu\nu} X_{\mu\nu}$, if the scalar $\Sigma$
transforms under the adjoint representation and develops a vev
\cite{ArkaniHamed:2008qp}.
Such operators have to be generated via loops involving particles of
mass $\Lambda$.

The non-Abelian group could be either the dark sector gauge group or a
group containing $U(1)_Y$.  We will focus on the latter option, as it
allows for grand unification and implies a simpler dark sector, and
will briefly return to the former option afterwards.

\subsubsection{Adjoint Scalar}
\begin{figure}[ht]
\centering
\includegraphics{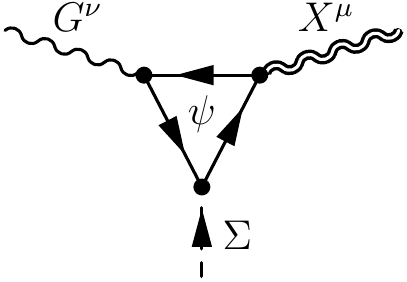}
\caption{The generation of an effective operator at the 1-loop level involving an adjoint scalar $\Sigma$ and a vector-like fermion $\psi$ 
that leads to kinetic mixing.}
\label{fig:SU5Adjoint}
\end{figure}

Consider first a dark $U(1)_X$ and a visible sector with a GUT
gauge group $G \supset U(1)_Y$, whose gauge bosons are denoted by
$G^\mu$.  We introduce a scalar $\Sigma$ that transforms under the
adjoint representation of the non-Abelian group and is uncharged
under $U(1)_X$.  In addition, we introduce a vector-like fermion $\psi$
with mass~$\Lambda$ that transforms non-trivially under both $G$ and
$U(1)_X$.  Then the diagram in \Figref{fig:SU5Adjoint} generates the
effective operator $\frac{1}{\Lambda} \Sigma \, G^{\mu\nu} X_{\mu\nu}$.
This diagram can be drawn for any group~$G$ and any (non-singlet)
representation of $\psi$, since the coupling of $\psi$ to the adjoint
scalar is the same as the coupling to the gauge bosons of $G$ (up to a
factor of $\gamma^\mu$).  

Once $\Sigma$ develops a vev $\braket{\Sigma}$ (chosen such that the SM
gauge group remains unbroken) we obtain kinetic mixing between $B^\mu$
and $X^\mu$.  Assuming that $\braket{\Sigma}$ is also responsible for the
breaking of the GUT group, the vev $\braket{\Sigma}$ is of order the unification scale $\MU$,
leading to the estimate
\begin{equation} \label{eq:SU5-24}
	\epsilon \sim \frac{g g_X y_\Sigma}{16\pi^2} \frac{\braket{\Sigma}}{\Lambda} \sim
	\frac{g g_X y_\Sigma}{16\pi^2} \frac{\MU}{\Lambda} \gtrsim
	\frac{y_\Sigma}{16\pi^2} \frac{\MU}{\MP}
	\sim 10^{-4} \, y_\Sigma \, ,
\end{equation}
for $\mathcal{O}(1)$ gauge couplings, 
where $g$ is the GUT gauge coupling and
$y_\Sigma$ is the coupling of $\psi$ 
to $\Sigma$.
Thus, to satisfy experimental bounds additional
suppression is required and can be obtained most easily by setting
the Yukawa coupling $y_\Sigma$ to a sufficiently small value.

\subsubsection{Fundamental and Other Representations}
\begin{figure}[ht]
\centering
\includegraphics{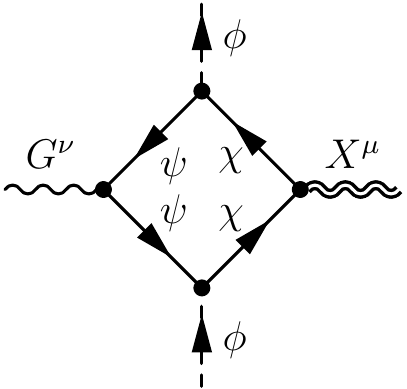}
\caption{The generation of an effective operator leading to kinetic mixing,
utilizing vector-like fermions $\psi$ and $\chi$ as well as a scalar
$\phi$ transforming under a representation different from the adjoint.
}
\label{fig:SU5Fundamental}
\end{figure}

Using a scalar $\phi$ transforming under a representation different from
the adjoint, we can generate the effective operator
$\frac{1}{\Lambda^2} \phi^\dagger G^{\mu\nu} \phi \, X_{\mu\nu}$ via the
diagram in \Figref{fig:SU5Fundamental}.
If the unified group is broken by an adjoint vev, the contribution from
$\phi$ will be subdominant compared to the one from the adjoint unless 
$y_\Sigma \lesssim y_\phi^2
\frac{\braket{\phi}^2}{\Lambda\braket{\Sigma}}$.
Let us explore the possibilities arising in this case.
Of course, there are many
possible choices, but not every possibility that is allowed by group
theory is phenomenologically viable.

For example, consider $G=SU(5)$ with a fundamental scalar
$\phi \sim (\mathbf{5},0)$, where the numbers in parentheses indicate the $SU(5)$ representation
and the $U(1)_X$ charge.
Then the diagram in
\Figref{fig:SU5Fundamental} can be realized, for instance, with the
vector-like fermions $\psi \sim (\mathbf{5},q_\psi)$ and
$\chi \sim (\mathbf{10},q_\psi)$.%
\footnote{The diagram can be drawn with different $U(1)_X$ charge
assignments $q_\chi \neq q_\psi$ as well, but then $\phi$ needs a
non-zero charge, which implies $\braket{\phi} \lesssim m_X$ and makes it
impossible to obtain observable kinetic mixing.}
However, as $\phi \sim (\mathbf{3},\mathbf{1}) + (\mathbf{1},\mathbf{2})$
under $SU(3)_\text{c} \times SU(2)_\text{L}$, the vev of $\phi$ can only
be non-zero for the electrically neutral component, the analog
of the SM neutrino in the fermionic $\overline{\mathbf{5}}$ multiplet.
This component couples to both $B^{\mu\nu}$ and $W_3^{\mu\nu}$ in such a
way that after electroweak symmetry breaking it has a non-zero coupling
only to $Z^{\mu\nu}$ but not to $F^{\mu\nu}$.  Consequently, this case
is not interesting for us, since it does not lead to kinetic mixing of
$X^\mu$ with the photon.

Moreover, $\phi \sim \mathbf{5}$ of $SU(5)$ cannot have a GUT-scale vev
since all its components are charged under the SM gauge group.  With an
electroweak-scale vev and $\Lambda \sim \MU$, the contribution to any
kinetic mixing is suppressed by
$(\frac{v_\text{EW}}{\MU})^2 \sim 10^{-28}$ and thus much smaller than
the minimal contribution from gravity discussed in \Secref{sec:Gravity}.
Thus, in order to obtain kinetic mixing of a relevant size in cases
involving a SM non-singlet scalar, we would have to lower $\Lambda$ much
below $\MU$.

As a consequence, we restrict our attention to scalar multiplets that
contain a SM singlet and can thus obtain a large vev yielding a sizable
$\epsilon$ even if $\Lambda \gtrsim \MU$.  Sticking to $SU(5)$, the
smallest viable multiplet is the $\mathbf{75}$.%
\footnote{We note that the $\mathbf{75}$ has been used instead of the adjoint
$\mathbf{24}$ to break $SU(5)$ in the missing partner mechanism to solve the doublet-triplet problem \cite{Masiero:1982fe}.}
Then the smallest fermion
multiplet we can use is $\psi=\chi \sim (\mathbf{10},q_\psi)$.  Giving a
vev (only) to the SM-singlet component of~$\phi$, the only non-zero term
in the decomposition of $\braket{\phi^\dagger} G^{\mu\nu} \braket{\phi}$
is the one containing $G_{24}^{\mu\nu}=B^{\mu\nu}$.  Hence, we generate
kinetic mixing with $B^\mu$ (but not $W_3^\mu$) and thus with both the
photon and the $Z$, as desired.  Its size is of order
\begin{equation} \label{eq:SU5-75}
	\epsilon \sim \frac{g g_X y_\phi^2}{16\pi^2} \frac{\braket{\phi}^2}{\Lambda^2} \sim
	\frac{g g_X y_\phi^2}{16\pi^2} \left( \frac{\MU}{\Lambda} \right)^2 \gtrsim
	\frac{y_\phi^2}{16\pi^2} \left( \frac{\MU}{\MP} \right)^2
	\sim 10^{-6} \, y_\phi^2 \,,
\end{equation}
for $\mathcal{O}(1)$ gauge couplings and $\braket{\phi}\sim\MU$, where now
$y_\phi$ is the coupling between $\phi$, $\psi$ and $\chi$.  As a result, an additional suppression
by one or two orders of magnitude due to small couplings or a smaller value of
$\braket{\phi}$ is sufficient to satisfy the bounds for
$m_X \lesssim 10^{-4}\eV$ or $m_X \gtrsim 1\MeV$. 

In order to give an example with a different unified group as well, let
us take $G=SO(10)$.  Then two simple possibilities to realize the diagram
of \Figref{fig:SU5Fundamental} are
$\phi \sim (\mathbf{126},0)$,
 $\psi \sim (\overline{\mathbf{16}},q_\psi)$,
 $\chi \sim (\mathbf{16},q_\psi)$, and
$\phi \sim (\mathbf{16},0)$, $\psi \sim (\mathbf{16},q_\psi)$,
 $\chi \sim (\mathbf{10},q_\psi)$.
These cases also offer the option of using fermions in the loop
that receive masses $\Lambda \sim \MU$ via couplings to additional
scalars transforming under $\mathbf{45}$, $\mathbf{54}$ or $\mathbf{210}$
and developing GUT-scale vevs to break $SO(10)$.%
\footnote{Assuming the vector-like masses that are independent of GUT
breaking are subdominant.}
In this line of thought, $\phi \sim \mathbf{126}$ may be especially
interesting if it obtains a vev of order $10^{10}\GeV$ or larger that
also gives a mass to the right-handed neutrinos in the fermionic
$\mathbf{16}$.  According to \Eqref{eq:SU5-75},
$\braket{\phi}\sim10^{10}\GeV$ and $\Lambda\sim\MU$ would result in
$\epsilon \sim 10^{-14}$ for $\mathcal{O}(1)$ couplings.

\subsubsection{Non-Abelian Dark Sector}
If the gauge group in the dark sector is non-Abelian, we can obtain
kinetic mixing with the SM gauge boson $B^\mu$ in the same way
as for a non-Abelian visible sector.  Now the scalars have to
be charged under the dark gauge group.  If their vevs
$\braket{\Sigma}$ and $\braket{\phi}$ give a mass to the dark photon,
they are of order $m_X/g_X$, which leads to
\begin{equation}
	\epsilon \sim
	\frac{g_X  g' y_\Sigma}{16\pi^2} \frac{\braket{\Sigma}}{\Lambda}
	\sim \frac{y_\Sigma}{16\pi^2} \frac{m_X}{\Lambda} \,,
	\label{eq:Dark-24}
\end{equation}
for the adjoint scalar case, and 
\begin{equation}
	\epsilon \sim
	\frac{g_X  g' y_\phi^2}{16\pi^2} \frac{\braket{\phi}^2}{\Lambda^2}
	\sim \frac{y_\phi^2}{16\pi^2 g_X} \frac{m_X^2}{\Lambda^2} \,,
	\label{eq:Dark-75}
\end{equation}
for the case of a scalar not transforming in the adjoint.
Now $\Lambda$ cannot be very large if we are
to obtain observable kinetic mixing.  However, $\Lambda$ has to be large
enough to hide the electrically charged fermions $\psi$ and $\chi$ from
detection.  For $\Lambda > 1\TeV$,%
\footnote{Indirect searches for new physics may well set a significantly
stronger limit, depending on details of the dark sector.}
\Eqref{eq:Dark-24} yields $m_X \gtrsim 10^{14}\epsilon \eV$ in the
adjoint case with $y_\Sigma \sim 1$.  For scalars transforming under different representations
and $y_\phi \sim 1$,
\Eqref{eq:Dark-75}  leads to
$m_X \gtrsim 10^{13}\sqrt{g_X\epsilon} \eV$, which allows us to approach
the parameter space interesting for fixed target experiments for
$\epsilon \sim 10^{-6}$ and $g_X \lesssim 10^{-3}$.

In order to obtain a wider range of viable parameters, we can use a
scalar that breaks the non-Abelian dark group to $U(1)_X$ at a
sufficiently high scale, thus decoupling the vev involved in kinetic
mixing from the dark photon mass.  The minimal possibility is
$SU(2)_X$ together with an adjoint scalar.
A scenario of this kind leading to $\braket{\Sigma} \sim 10^4\GeV$ and
$\Lambda \sim 10^{16}\GeV$, which corresponds to $\epsilon \sim 10^{-14}$
for $\mathcal{O}(1)$ couplings, was presented in~\cite{Acharya:2017kfi}.

Finally, we can combine the possibilities discussed in this section by
considering non-Abelian groups in both sectors.  That is, we assume the
overall gauge group $G \times G'$, where in the simplest scenario
$G \supset U(1)_Y$ and $G' \supset U(1)_X$ are broken by the vevs of the
adjoint scalars $\Sigma$ and $\Sigma'$, respectively.  In the presence
of a vector-like fermion of mass $\Lambda$ that is charged under both
groups, we obtain \cite{Goldberg:1986nk}
\begin{equation}
	\epsilon \sim \frac{g g_X y_\Sigma y_{\Sigma'}}{16\pi^2}
	\frac{\braket{\Sigma} \braket{\Sigma'}}{\Lambda^2} \,.
\end{equation}
As the unification scales in the two sectors are not related
in general,
$\braket{\Sigma'}$ can be much smaller than $\MU$, which yields very
small values of $\epsilon$ even if all gauge and Yukawa couplings are of
order~$1$.  For example, $\epsilon \sim 10^{-14}$ for $\Lambda \sim \MP$,
$\braket{\Sigma} \sim \MU$, and $\braket{\Sigma'} \sim 10^8\GeV$.

\section{Summary}
\label{summary}

Because simple dark matter candidates such as a fourth generation heavy neutrino with mass of order a few GeV,
or the lightest supersymmetric particle such as a neutralino
with mass of order a few hundred GeV, have been excluded (in the case of the former), and severely constrained (in the case of the latter), a plethora of dark matter candidates have arisen
with varying degrees of simplicity. Among these,
there are many theories with a presumed stable dark matter
candidate which has no SM gauge interactions, and 
instead carries a charge under some hidden sector gauge group
which is often assumed to be $U(1)_X$. This opens up the possibility that the gauge field associated with the hidden
$U(1)_X$, can have a kinetic mixing term with the SM photon.

There is, however, a large body of constraints 
on the mixing parameter $\epsilon$ which lead to 
upper limits of order $10^{-7}$ for a wide range of
dark photon masses
between $\mathcal{O}(10^{-14})\eV$ and $\mathcal{O}(100)\MeV$,
with significantly stronger bounds ($\epsilon < 10^{-15}$)
for dark photon masses around $1\keV$ as seen in \Figref{fig:Bounds}.

If there are fields which are charged under both the SM and 
the hidden $U(1)_X$, then one expects (barring a fine-tuning)
kinetic mixing at the one-loop level, with a value
given by the estimate in \Eqref{eq:Standard1L}, which is
not much smaller than $10^{-2}$ and
in rather severe disagreement with the experimental limits seen in \Figref{fig:Bounds}.

In this paper, we have considered both bottom-up and
top-down approaches to building a model with 
sufficiently small kinetic mixing. 
The bottom-up approach is necessarily complicated
by the fact that fields must be charged under only
a single $U(1)$, to avoid one-loop mixing. 
To this end, we have considered a model
based on the right-handed neutrino portal 
which involves both the SM Higgs and a
hidden sector Higgs $H_X$. When $H_X$
acquires a vev, we can construct a two-loop
diagram for mixing above and below the weak scale.
Since the kinetic mixing in this case is proportional to unknown SM and hidden Yukawa couplings, the mixing parameter can be tuned to 
very small values.

We have also argued that gravity alone can lead to kinetic mixing. Though this occurs at the six-loop
level, it provides us with a lower limit to 
$\epsilon$ which can be as large as $10^{-13}$
if the hidden sector Yukawa coupling is of order one and the charge conjugation symmetry is broken in the hidden sector.

We have also considered the construction of kinetic mixing in top-down
models where all gauge groups are unified into a single GUT\@. Once again,
the prime difficulty is finding matter representations which are not charged under both
the SM and hidden $U(1)_X$ gauge groups. Indeed,
for the commonly studied $SO(10)$ and $E_6$ GUT
gauge groups, we found no representations
which allow us to escape the estimate in \Eqref{eq:Standard1L}. However, in $E_7$,
which breaks to $SO(10) \times U(1)_A \times U(1)_B$, the entire SM
$\mathbf{16}$ which originates in a $\mathbf{27}$ of $E_7$ is uncharged
under one linear combination of the two $U(1)$'s. However,
the model must be complicated by choosing the 
Higgs $\mathbf{10}$ from a different $E_7$ representation, the smallest
being the $\mathbf{133}$.
Models in $E_8$ GUTs are also possible. 

Finally, we also considered models of the form
GUT$\times U(1)_X$. In this case, we 
require a higher-dimensional operator to provide the kinetic mixing. If that operator is mediated
by Planck-scale physics, we can expect
a suppression of order $\MU/\MP$ over
the one-loop estimate. Higher order suppressions
are possible if we employ larger representations 
to break the GUT (such as the $\mathbf{75}$ in the case of $SU(5)$).

Of course nature has already decided
if dark matter resides in a hidden sector and 
communicates with the visible sector through kinetic mixing. We rely on experimental discovery
to confirm or exclude this class of theories. 
We have seen, however, that the construction
of such theories, whether within the context of a GUT or not, is highly non-trivial. Furthermore, kinetic
mixing through
gravity may already preclude some range of dark photon masses.

\section*{Acknowledgements}

\noindent
We would like to thank Bohdan Grzadkowski and Jörg Jäckel for helpful discussions.
Special thanks are due to Javier Redondo for providing a Mathematica
notebook with experimental bounds on kinetic mixing that we used to
produce \Figref{fig:Bounds}.
The work of TG and KAO was
supported in part by DOE grant DE-SC0011842 at the University of
Minnesota.
JK acknowledges financial support from the Fine Theoretical Physics
Institute (FTPI) at the University of Minnesota and the Meltzer Research
Fund, and would like to thank the FTPI and the Abdus Salam ICTP (Trieste,
Italy), for their hospitality during the work on this project.

\bibliographystyle{utcaps}
\bibliography{KineticMixing}

\end{document}